\documentstyle[twocolumn,prl,aps,epsfig]{revtex}

\begin{document}
\draft
\input{psfig}
\title{Ordering of the lamellar phase under  a shear flow}
\author{F. Corberi}
\address{Istituto Nazionale per la Fisica della Materia,
Unit\`a di Salerno {\rm and} 
Dipartimento di Fisica, Universit\`a di Salerno,
84081 Baronissi (Salerno), Italy}

\author{G. Gonnella$^{(+)}$ and A. Lamura$^{(+)(*)}$}
\address{$^{(+)}$ Istituto Nazionale per la  Fisica della Materia
{\rm and} Dipartimento di Fisica, Universit\`a di Bari {\rm and}
Istituto Nazionale di Fisica Nucleare, Sezione di Bari, via Amendola
173, 70126 Bari, Italy\\
$^{(*)}$ Istituto Applicazioni Calcolo, CNR, Sezione di Bari, via Amendola 122/I 
- 70126 Bari - Italy}
\date{\today}
\maketitle
\begin{abstract}
The dynamics of a system
quenched  into a state with  lamellar order 
and subject to an uniform shear flow
is solved  in the large-$N$ limit.
The description is based on the  Brazovskii free-energy and 
the evolution 
follows  a convection-diffusion equation.
 Lamellae order preferentially 
 with the normal along the vorticity direction.
Typical lengths grow as $\gamma t^{5/4}$ (with logarithmic corrections)
 in the flow direction
and logarithmically  in the shear direction.
Dynamical scaling holds in the two-dimensional case while
it is violated in $D=3$.

\end{abstract}


Many systems in nature exhibit lamellar order.
One example is a diblock copolymer melt where
chains of type A and B covalently bonded end-to-end
in pairs segregate  at low temperatures
with A-B junctions forming a stack of lamellae \cite{Bat}.
In ternary mixtures a lamellar phase is stable
with ordered sheets of surfactant 
separating  (say) oil and water domains \cite{GS94}.
Lamellar order is also observed  in
 Raleigh-B\'enard cells, where 
convective rolls form
above the convective threshold \cite{CH}. 
Further examples include smectic liquid crystals \cite{Deg},
dipolar fluids with long-range interactions \cite{SD}
 and chemically reactive binary mixtures \cite{GC}.
A theoretical model for the general description of the 
lamellar-disordered phase transition was proposed 
by Brazovskii \cite{bra} 
who showed  the first-order character of
the  transition induced by fluctuations.

Lamellar phases under an applied shear flow show a very rich  
behavior which is relevant for many applications~\cite{larson}.
A  variety of transitions in morphology and orientation
occur  as shear rate and temperature are changed.
Stable configurations of lamellae 
lying along the flow with the normals differently oriented
 have been observed~\cite{Kop,Win}
and analyzed evaluating  the effects of the flow on the fluctuation 
spectra~\cite{cates,Fr95,MG,Fr}.

Non-stationary properties are far less considered \cite{KD96,Ba98,DCV,GIC}.
In this Letter we study 
the effects of  a simple planar  shear flow on the ordering 
 of the  lamellar phase in a system 
quenched from an initially  high temperature disordered state.
We present the first analytical 
results 
on  the kinetics  of this system by solving the Brazovskii
model in the limit of an 
infinite number of components of the order parameter.
This  is one of the few methods allowing an explicit
solution for phase ordering systems~\cite{B94}.
A similar approach for fluids under  shear flow
has been used in~\cite{OK,Fre,FL,O87,daw,HM,BR,CGL}.

The behavior of  quenched  binary mixtures without imposed flows 
is characterized by  dynamical scaling: The structure factor obeys
$ C(\vec k, t) = R(t)^D f[k R(t)]$  
where $R(t) \sim t^{\alpha}$ is the typical domain size
 and $D$ is the space dimensionality~\cite{Gunton,B94}.
In the case of  a fluid with lamellar order, 
if the order parameter is not conserved as 
in the Swift-Hohenberg model for 
 Raleigh-B\'enard convection \cite{SH}, regimes exhibiting  dynamical 
scaling have been found~\cite{BV}.
In  models with conserved order parameter,
when diffusion is the only segregating physical mechanism,
 simulations show  the entanglement of the fluid 
into  frozen intertwined structures  \cite{LG92,BO90}.
In this case   the effects of hydrodynamical modes are  crucial 
for reaching order on large scales \cite{yeo}.

When shear is applied, our results 
show that  lamellae grow  preferentially
with the perpendicular orientation~\cite{Fr95}, 
namely along the plane 
formed by the flow and the shear (velocity gradient) 
directions.    
 Their typical size, obtained from 
 the second  momentum
of the structure factor, grows as  $\gamma t^{5/4} \sqrt{\ln t}$ in the flow
direction and as  $ \sqrt{\ln t}$ in the shear  direction.
Surprisingly we find that
dynamical scaling is obeyed in two dimensions but not in $D=3$.
Our results concern
the cases  of conserved and not conserved order parameter, 
and  apply to most of the systems mentioned above.

We consider the Brazovskii free energy
\begin{equation}
{\cal F}\{\phi \}=\int d\vec r \left \{ \frac {r}{2} \phi ^{2} +
\frac {u}{4} \phi ^{4} - \frac {b} {2} \vert \nabla \phi \vert ^{2}
+ \frac {1} {2} ( \nabla ^{2}  \phi)^{2}  \right \}
\label {freen}
\end{equation}
where $\phi(\vec r,t)  $ is the order parameter field and $u  > 0$.
With a negative value of  $r$
and $b < 0$ the  system orders
 in one of the two minima of the local potential.
However, 
when  $b > 0 $ interfaces are favoured and   a modulated state with
wavevector $ k_M = \sqrt{b/2}$ is  stable~\cite{bra}.

When a flow is imposed, the kinetics  
 can be described by the convection-diffusion equation \cite{On97}
\begin{equation}
\frac {\partial \phi}{\partial t}+\vec \nabla \cdot (\phi \vec v)
=- \Gamma_p \frac {\delta {\cal F}} {\delta \phi} \quad, \hskip1cm p=0,2  
\label{motion}
\end{equation}
where
\begin{equation}
\Gamma_0 = \Gamma, \qquad
\Gamma_2 = - \Gamma \nabla ^{2}
\label{c}
\end{equation}
and $\Gamma$ is a mobility coefficient.
 $p=0 $ describes  systems with non-conserved order parameter 
(NCOP) and
corresponds to the Swift-Hohenberg equation \cite{SH};
the case $p=2$ is for conserved order parameter (COP)
 and applies for example to copolymer melts.
For  uniform shear flow in the $x$-direction 
 $ v_x =\gamma~y$,  $ \gamma$ being the shear rate.
Eq.(\ref{motion}) neglects thermal fluctuations and possible effects due to 
differences in viscosities between the two components~\cite{Fr95}.
The complete description should  
take into account  the coupling of
Eq.~(\ref{motion}) with the Navier-Stokes equation. However,  
the study  of Eq.~(\ref{motion}) is a prerequisite for any
more general theory and usually  applies to intermediate temporal regimes,
as recognized for phase separation of 
simple binary mixtures~\cite{B94}. 

Eq.~(\ref{motion}) 
can be studied analytically generalizing the field $\phi$
to a vector order parameter with $N$ components and taking the
large-$N$ limit \cite{MZ85}.
One obtains
the following equation in Fourier space for the structure factor $C(\vec k,t)=
\langle \phi (\vec k,t) \phi (- \vec k,t)\rangle $
\begin{eqnarray}
\frac {\partial C(\vec k,t)} {\partial t} &-& \gamma k_x 
\frac {\partial C(\vec k,t)} {\partial k_y}= \nonumber \\
&-& 2 \Gamma k^{p}\left [
r+u S(t)-bk^{2}+k^{4} \right ] C(\vec k,t)
\label{motoc}
\end{eqnarray}
where we have dropped the component indices due to internal
symmetry. $S(t)$ has to  be computed self-consistently through
\begin{equation}
S(t)=\int _{\vert k \vert < \Lambda} \frac {d\vec k}{(2\pi)^D} C(\vec k,t),
\label{self}
\end{equation}
$\Lambda$ being an ultraviolet cut-off.
For a symmetric mixture  
quenched from a high temperature homogeneous phase an appropriate initial 
condition is $C(\vec k,0)=\Delta$,  where $\Delta$ is a constant.
Eq.~(\ref{motoc}) can be integrated yielding 
\begin{equation}
C(\vec k,t)=\Delta e^{-2\Gamma\int _0 ^t d\tau \vec {\cal K}^p(\tau)\left [ {\cal K}^4(\tau) 
-b {\cal K}^2(\tau)+r+u S(t-\tau) \right ]}
\label{cc}
\end{equation}
where $ \vec {\cal K}(\tau)\equiv \vec k+\gamma \tau k_x \vec e_y$.
Defining
 \begin{equation}
Q(t) \equiv \int_0^t d\tau [r + u S(t - \tau)]
\label{q}
\end{equation}
 the analysis can be carried out asymptotically 
through  the  {\it ansatz} 
\begin{equation}
Q(t) = k_M^{4} t -  \frac {1} {2 \Gamma k_M^p }
(2 \ln t -  \ln \ln t  + \ln v_p),
\label{ansatz}
\end{equation}
where  $v_p$ is a constant to be determined. 
Eq.(\ref{ansatz}) will be  
justified {\it a posteriori} by proving 
the solution of 
the self-consistency problem. Specifically, the large time 
 value 
$ S_{\infty}$ calculated through (\ref{self}),
inserted into (\ref{q}), must  satisfy (\ref{ansatz}) asymptotically, namely
\begin{equation}
S_{\infty} =   (k_M^{4}- r )/u
\label{sinfty}
\end{equation}
We now  briefly illustrate  this for  COP in $ D = 3 $~\cite{nota1}
setting $\Gamma = 1 = u = b = -r $.

\begin{figure}
\epsfig{file=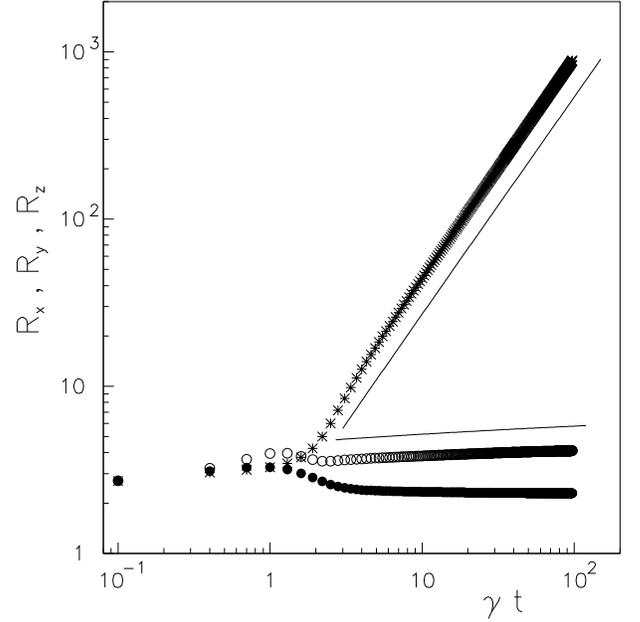,bbllx=45pt,bblly=159pt,bburx=526pt,bbury=647pt,width=8.2cm,clip=}
\caption{The typical lengths as a function of the shear strain 
in the $x$ ($\ast$), $y$ ($\circ$) and $z$ ($\bullet$) directions.
The  lines are proportional to  $\gamma t^{5/4} (\log t)^{1/2}$ and
$(\log t)^{1/2}$.} 
\end{figure} 

The symmetry of the problem suggests the use of the cylindrical set of 
variables 
$(k_x,k_\perp,\theta)$, where $k_y=k_\perp \cos \theta$ and 
$k_z= k_\perp \sin \theta$.
Naive power counting applied to Eq.~(\ref{cc}) indicates that 
wave vectors in the flow direction scale asymptotically as 
$k_x\sim t^{-3/2}$ while  in the perpendicular directions
$k_\perp - k_M \sim t^{-1/2}$.
Then, introducing the new variables $X=k_x L_\parallel(t)$, 
$Q=(k_\perp-k_M)L_\perp (t)$, with 
$L_\parallel(t)=2\sqrt {2/3} k_M^2 \gamma t^{3/2}$
and $L_\perp(t)=2\sqrt{2t}k_M^2$, 
and taking into account Eq.~(\ref{ansatz}),  Eq.~(\ref{cc}) reads  as 
\begin{equation}
C(\vec k, t) =  v_2 \Delta  
 \frac{t^2}{\ln t} e^{-[f(X,Q,\theta) 
+ g(X,Q,\theta ,t)]}
\label{cdikappa}
\end{equation}
with
\begin{equation}
f = Q^2+\sqrt{3}Q X\cos \theta +X^2 \cos ^2 \theta 
\label{cscal}
\end{equation}
and $g(X,Q,\theta,t)= \sum_{j=1}^{16} t^{-j/2} g_j$, where  
$g_j$ are polynomials in   $X,Q,\theta, \log t$. For long times
$g$  can be shown to provide a relevant  contribution
$g_2 \simeq \frac {9}{160} \frac{X^4}{k_M^6} - \frac {3}{4 k_M^6} X^2 \ln t$
only  in the region  $Q\sim 0, \cos \theta \sim 0$.
Then  the  integration over $Q$ and $\theta$ can be performed yielding
\begin{equation}
S(t) \simeq \frac {\sqrt 3 \Delta }{ 32  k_M^{3}\pi ^{\frac{3}{2}}}
 \frac {v_2}{\gamma \ln t}
 \int _{-\infty} ^\infty 
dX e^{-\frac {X^2}{8} - \frac{g_2}{t}  }
I_0({X^2}/{8}) 
\label{integral}
\end{equation}
where $ I_0(z)$ is a Bessel function.
The  integral of Eq.~(\ref{integral}), 
evaluated by 
the asymptotic expansion of $I_0(z)$ at large $z$,  
 behaves as  $ \ln t /\sqrt{\pi}$. 
Hence at large times $S(t)$ approaches the constant  $ S_{\infty}
 = \frac{\sqrt{3} v_2 \Delta } 
{32 \pi^2 k_M^3 \gamma} $. 
A comparison with Eq.(\ref{sinfty}) fixes  
$v_2 = \frac {32 \pi^2 k_M^3 \gamma (k^4_M + 1)}{
\sqrt{3} \Delta }$  and verifies the ansazt.

Due to the presence of $g$,  
the function $C(\vec{k},t)$ of Eq.(\ref{cdikappa})
cannot be cast in a scaling form.
However, different scaling behaviors are obeyed in the regions
where the functions $f$ or $g$ can be respectively neglected.
In particular, for $Q\sim 0, \cos \theta \sim 0$, where $g$ dominates,
$C(\vec{k},t)$ scales  in the $x$-direction with respect to the  length  
$l_\parallel(t) = (2/5)^{1/4}\sqrt{k_M} \gamma t^{5/4} $
 different from $L_\parallel(t)$.
In $D=2$, on the other hand,  where the contribution
of the function $g$ is always negligible,  
one finds the scaling form
$ C(\vec k, t) =  v_2 \Delta 
 t^2  e^{-[f(X,Q,0) ]}$.

\begin{figure}
\epsfig{file=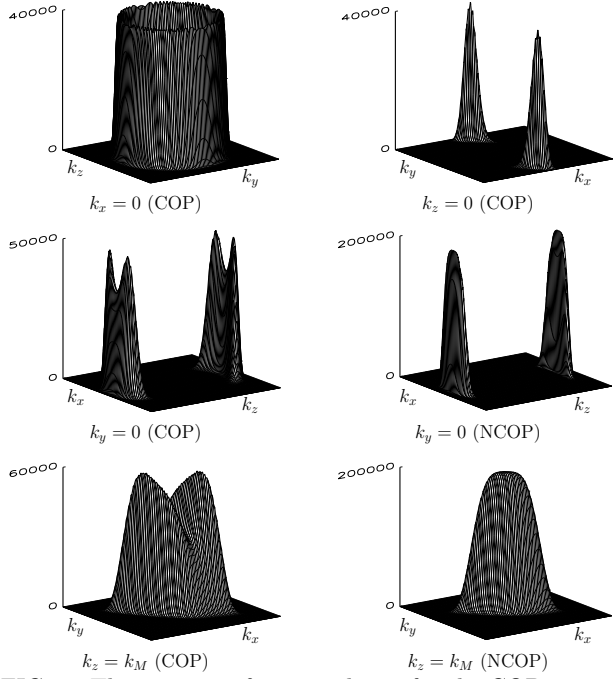,bbllx=80pt,bblly=235pt,bburx=550pt,bbury=755pt,width=8.2cm,clip=}
\caption{The structure factor 
is shown for the COP case on the planes $k_x=0$, $k_y=0$, $k_z=0$,
$k_z=k_M$ and for the NCOP case on the planes $k_y=0$, $k_z=k_M$.}
\end{figure}

Ordering properties are usually inferred from the 
momenta of the structure factor. 
We define 
$R_x =  [\int 
{d\vec k} C(\vec k,t) /  \int 
{d\vec k} k_x^2 C(\vec k,t)]^{1/2}
$ and similarly for the shear and vorticity  directions. 
In the absence of dynamical scaling 
 definitions of $R_x$ based on different 
momenta of $C(\vec{k},t) $ may lead to different results. However in this case
it can be proven that changing the order of the momentum does not change
the growth exponent but only the logarithmic correction. 
>From 
Eq.~(\ref{cdikappa}) 
we find  $R_x \sim \gamma t^{5/4} \sqrt{\ln t}$,
$R_y  \sim  \sqrt{\ln t}$ and 
$R_z \sim  k_M^{-1}$.
The behavior of $R_x$ shows the relevance of the scaling with respect to
$l_\parallel(t)$ for the ordering of the system. 
The growth of $R_y$  
indicates  that lamellae order preferentially in the plane $x-y$.
In $D=2$ we find  $R_x \sim \gamma t^{3/2}$ 
and $R_y \sim  k_M^{-1}$. 
The same results are  obtained with non-conserved order parameter.
 The behavior of $R_x, R_y, R_z$  resulting from the  
 numerical integration of Eq.~(\ref{motoc}) 
with an adaptive grid algorithm is  shown in Fig.~1. 
After the  initial isotropic evolution,
the shear-induced anisotropy  becomes evident  for values
of the strain  $\gamma t$  larger than one and agreement with the
analytical behavior is observed. 

It is also useful to  illustrate  the behavior of the structure factor
(\ref{cdikappa}) shown in Fig.~2.
 The maxima of $C(\vec{k},t)$ are located at
$ k_x^2 = 5 \frac {1}{k_M^4 \gamma^2} \frac{\ln t}{t^3}, k_y=0, 
k_z^2=k_M^2$ for COP and at $ k_x=0, k_y^2 + k_z^2=k_M^2$ for NCOP.
The shape of $C(\vec{k},t) $ on the planes $k_x =0 $ and $k_z = 0 $
is qualitatively similar for COP and NCOP.
At  $k_x = 0 $, in particular, the wavevectors are not affected by the flow
and the structure factor has a circular structure with
radius $k_M $. 
At $ k_y =0 $  or $ k_z = k _M$ the behavior of $C(\vec{k},t)$ depends on the 
conservation law. The maxima of $C(\vec{k},t)$ 
developed with NCOP are splitted with COP
into a pair of symmetric narrowing peaks. This pattern is 
typical of the case with conserved dynamics and
has been observed in other segregating systems
under  shear flow~\cite{CGLE}.

\begin{figure}
\epsfig{file=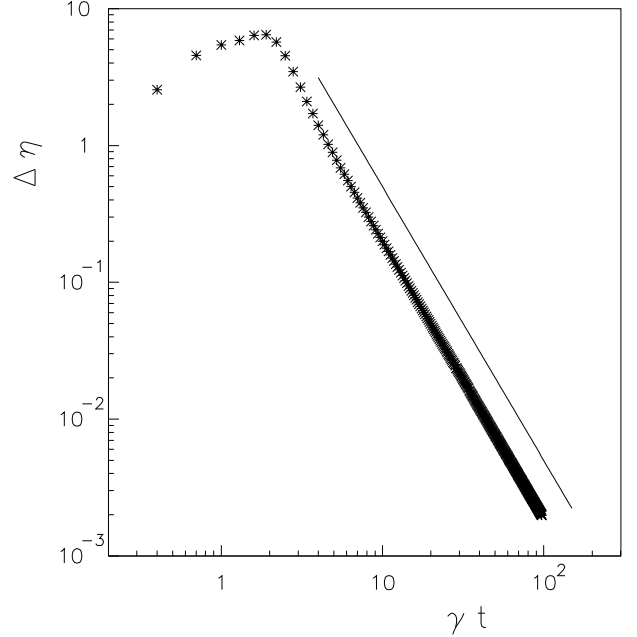,bbllx=38pt,bblly=145pt,bburx=523pt,bbury=660pt,width=8.2 
cm,clip=}
\caption{The excess viscosity as a function of the shear strain.
The straight line has slope -2.} 
\end{figure} 
Finally we turn to the study of the  rheological properties.
In phase separation of  binary mixtures the flow 
acts against the surface tension inducing  stretching of domains
followed by break-up processes and burst of small bubbles~\cite{OND,prl2}.
An excess viscosity   $\Delta \eta $ 
is measured~\cite{KSH}: it reaches 
a maximum generally for  $\gamma t \geq 1 $ 
and later decays.
In our case we  measure the evolution of the shear stress $\sigma_{xy}$
deducible in a general way from (\ref{freen}) in terms of the structure factor;
it is given by 
$\sigma_{x y}(t) = \int 
\frac {d\vec k}{(2\pi)^D} k_x k_y (2k^2-b)C(\vec k,t)$ \cite{daw}.
An excess viscosity can be introduced as 
$\Delta \eta (t)= - \sigma_{x y}(t)/\gamma$ \cite{Onu}.
Using Eq.~(\ref{cdikappa}) we find $ \Delta \eta \sim 
\gamma ^{-1}t^{-2}$. 
 The numerical results of 
 Fig.~3  show that, after reaching a maximum, 
the decay of $\Delta \eta$ agrees with the analytical behavior.
The other rheological indicators can be shown to behave  similarly.

In conclusion we have studied the ordering kinetics of
a lamellar phase in shear flow by solving  the dynamics 
of the convective-diffusion equation in the large-N limit.
Regarding the debated question of the stable orientation,
our results are in agreement with the expectation 
of the stability of the perpendicular phase at high shear rates~\cite{Fr95}.
In   $D=3$ 
a logarithmic growth law is found along the shear direction
while $R_x\sim \gamma   t^{5/4}\sqrt {\ln t}$. 
Interestingly, the same exponent $5/4$ is found for simple
binary fluids~\cite{BR,CGL}. 
Our results show that the scaling depends
on dimensionality and is violated in  $D=3$.
This is at variance with the case without shear where
the same approach gives a scaling form independently of $D$~\cite{fu}.
Violation of scaling is quite uncommon~\cite{CZ,B94}. 
In this context this phenomenon
may be related to the existence of two lengths $L_\parallel $, $l_\parallel$
growing with different exponents. 
 
A natural question is the relevance of the behavior of the large-N model 
to the physical case with $N=1$. 
As  without flow, symmetry 
arguments  suggest  an analogy between
the coarsening behavior of lamellar phases and that of
vectorial models and  one would argue that the large-N 
results give reliable indications  also for the physical case~\cite{maz};
in particular the large-N exponent is the one expected 
 asymptotically \cite{BV}. 
 Simulations of the scalar case with shear
could  elucidate this point; 
however 
 strong 
finite size effects make 
the evaluation of the growth exponent difficult
(see \cite{prl2} and refs. therein). 
We also mention that it will be important to study how 
the presence of hydrodynamics affects the picture provided in this Letter.

~\\
We thank Mario Pellicoro and  Marco Zannetti for useful discussions. 
F.C. and G.G. acknowledge M.Cirillo and R. Del Sole 
for hospitality in Rome University and 
 support by the  PRA-HOP 1999 INFM and MURST (PRIN 1999).


\begin{references}
\bibitem{nota1}
The proof is similar for NCOP; 
details  will appear elsewhere.
\end{references}
\end{document}